\documentclass[ preprint, aps, prl, one column, showkeys, groupedaddress]{revtex4-1}
\usepackage{graphicx,epsfig}
\usepackage{amssymb}
\usepackage{amsmath}
\usepackage{bm}
\usepackage{textcomp}
\usepackage{soul,xcolor}

\usepackage{color}

\begin{document}
\setcounter{page}{1}
\setstcolor{red}

\title[]{Ultra-high quality two-dimensional electron systems}
\author{Yoon Jang \surname{Chung}\textsuperscript{$\dag$}}
\author{K. A. \surname{Villegas-Rosales}\textsuperscript{$\dag$}}
\author{K. W. \surname{Baldwin}}
\author{P. T. \surname{Madathil}}
\author{K. W. \surname{West}}
\author{M. \surname{Shayegan}}
\author{L. N. \surname{Pfeiffer}}
\affiliation{Department of Electrical Engineering, Princeton University, Princeton, NJ 08544, USA  }
\date{\today}

\begin{abstract}

Two-dimensional electrons confined to GaAs quantum wells are hallmark platforms for probing electron-electron interaction. Many key observations have been made in these systems as sample quality improved over the years. Here, we present a breakthrough in sample quality via source-material purification and innovation in GaAs molecular beam epitaxy vacuum chamber design. Our samples display an ultra-high mobility of $44\times10^6$ cm$^2$/Vs at an electron density of $2.0\times10^{11}$ /cm$^2$. These results imply only 1 residual impurity for every $10^{10}$ Ga/As atoms. The impact of such low impurity concentration is manifold. Robust stripe/bubble phases are observed, and several new fractional quantum Hall states emerge. Furthermore, the activation gap of the $\nu=5/2$ state, which is widely believed to be non-Abelian and of potential use for topological quantum computing, reaches $\Delta\simeq820$ mK. We expect that our results will stimulate further research on interaction-driven physics in a two-dimensional setting and significantly advance the field.


\end{abstract}
\maketitle

	Single crystal GaAs thin-film structures grown by molecular beam epitaxy (MBE) are often considered to be among the purest materials that can be made in the laboratory. Being nearly defect free, these structures provide an exceptional platform for exploring a diverse range of physical sciences, with extensive electronic and photonic applications. The highlight of ultra-high-quality GaAs films, however, is their utilization in the investigation of electron-electron interaction phenomena. Typically this is achieved by studying the low-temperature magnetotransport of two-dimensional electron systems (2DESs) hosted in modulation-doped GaAs quantum wells, where the electrons are spatially separated from the dopants to reduce electron-impurity scattering. A magnetic field applied perpendicular to the 2DES enhances the relative scale of Coulomb energy in the system by quenching the Fermi energy via Landau quantization. A plethora of many-body phases have materialized in GaAs 2DESs using this framework; the discovery of the odd- \cite{Tsui} and even-denominator \cite{Willett} fractional quantum Hall (FQH) effects as well as the observation of Wigner solid \cite{Wigner0,Wigner1,Wigner2,Wigner3}, stripe/nematic \cite{stripe1,stripe2}, and bubble phases \cite{Bubble} are some notable examples (for reviews, see, \cite{rev1,JainBook}).
	
	Naturally, these fascinating phases only started to emerge in experiments as sample quality increasingly improved. For example, after the first observation of the FQH effect \cite{Tsui}, it took about two decades of growth condition improvements to realize stripe/nematic phases in GaAs 2DESs \cite{stripe1,stripe2}. These developments continue to motivate the community to search for methods to produce ever-better-quality samples, as it is stimulating to anticipate what other physics is yet to be uncovered. Indeed, these efforts have extended to several other material systems with varying electronic properties, and numerous many-body phases have been observed in the magnetotransport/capacitance features of 2DESs hosted in AlAs \cite{Lay.APL,Ettiene.APL,Shayegan.Review,Bishop,Chung.PRM,Shafayat.PRL}, Si \cite{Lai.PRL.2004,Kott.PRB.2014}, graphene \cite{Du,Kim,Dean,Zibrov,Graphene.FQHE.2018}, ZnO \cite{oxide1,oxide2}, Ge \cite{Ge1,Ge2}, and WSe$_2$ \cite{WSe2}. While each and every one of these systems provide the opportunity to study exquisite interaction-driven physics in different perspectives, GaAs-based 2DESs remain important; see, e.g., \cite{Friess.Nat.Phys.2017, Hossain.Nat.Phys.2020} 
	
	
	A useful metric to quantify the quality of a 2DES is the electron mobility because it is straightforward to measure and is inversely proportional to the average electron scattering rate in the 2DES. High mobility values in 2DESs imply that electrons are less likely to scatter over prolonged trajectories in such samples, making them valuable platforms to study delicate many-body electron phases as well as ballistic or phase-coherent transport. After decades of development, electron mobility values as high as $\mu\simeq35\times10^6$ cm$^2$/Vs have been observed in GaAs 2DESs with densities near $n\simeq3\times10^{11}$ /cm$^2$ \cite{PhysicaE,Umansky,mobilityrocks,Ga1}. However, despite subsequent efforts, the mobility of modern state-of-the-art GaAs 2DESs has been in a stalemate for more than a decade now. Here we present a breakthrough in the MBE crystal growth of ultra-high-quality GaAs 2DESs that enhances the mobility to $\mu\simeq44\times10^6$ cm$^2$/Vs at electron densities as low as $n\simeq2\times10^{11}$ /cm$^2$. In the low-density ($n\lesssim1.5\times10^{11}$ /cm$^2$) regime, where scattering by \textit{residual} (background) impurities is dominant, our samples show twice the mobility of previous samples \cite{PhysicaE,Umansky,mobilityrocks,Ga1}. Low-temperature magnetotransport traces taken in a sample of this class show new FQH states, indicating great prospect for future studies of interaction-driven physics in the two-dimensional setting.

\begin{figure*} [t]
\centering
    \includegraphics[width=.9\textwidth]{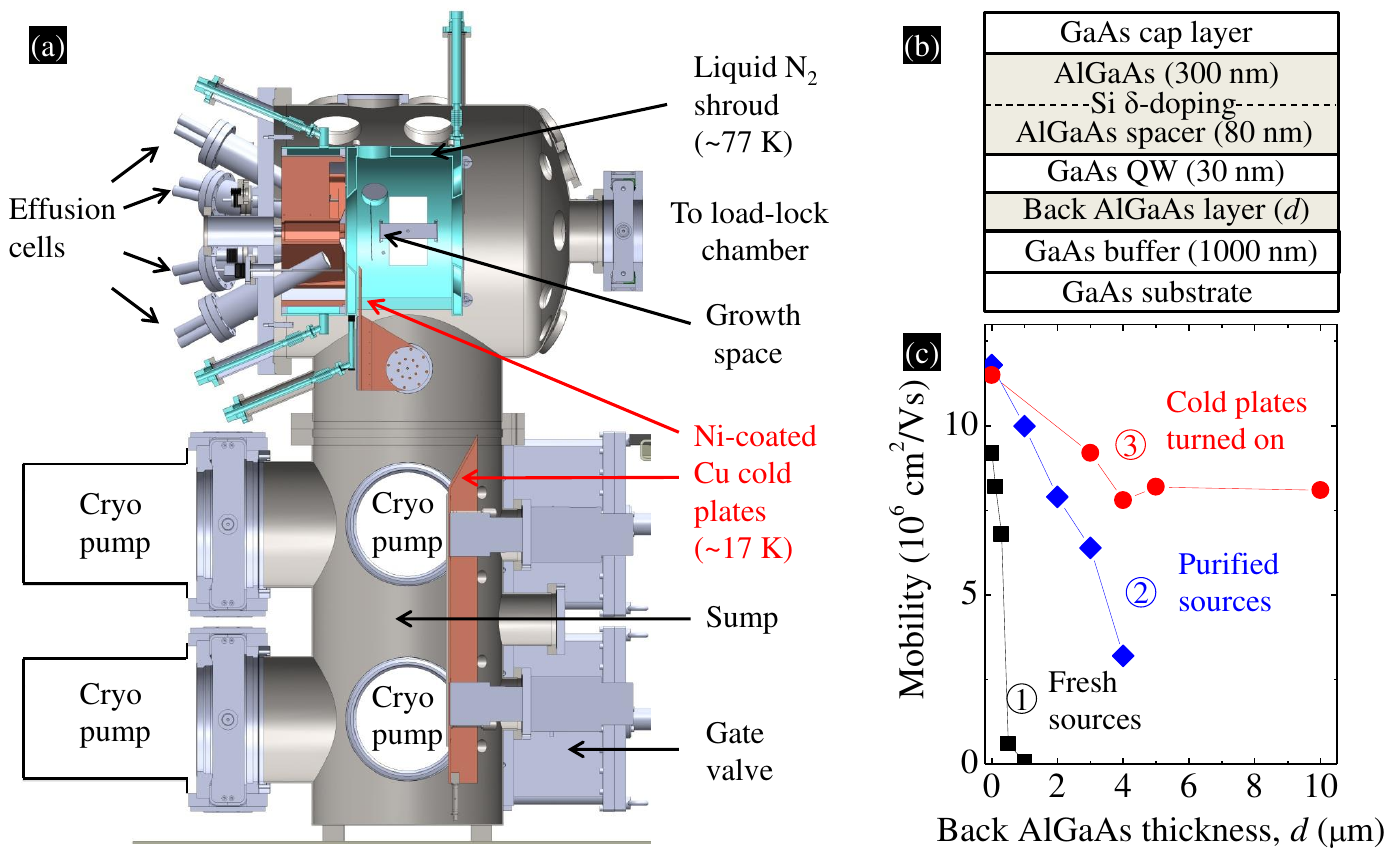}

\caption{\label{fig1} Improving vacuum quality and its assessment in a state-of-the-art MBE chamber. (a) Schematic diagram of the MBE chamber used in this study. In addition to four large (3000 $l$/s) cryopumps operating at $\sim10$ K, there are three auxiliary cryo-cooled, Cu cold plates ($\sim17$ K on their back side) that pump the chamber. One $\sim5\times5$ inch$^2$ cold plate is located in close proximity to the growth space while two $\sim12\times15$ inch$^2$ cold plates are in the sump. (b) Sample structure used to analyze vacuum quality. If the source material is pure enough, the surface-segregation of impurities during the growth of the back AlGaAs layer can be exploited to evaluate the vacuum quality by analyzing the mobility measured in the GaAs quantum well. (c) Mobility of samples that have the structure shown in (b), with varying back AlGaAs thicknesses ($d$) and vacuum/source purity conditions. The 2DES density is $n\simeq2.0\times10^{11}$ /cm$^2$. The data sets for different conditions are color coded, where the data shown in black, blue, and red were taken from samples grown in that chronological order; they correspond to samples grown with: (1) a fresh batch of sources, (2) purified sources, and (3) purified sources and the Cu cold plates turned on.}
\end{figure*}

	Numerical calculations have suggested that residual impurities are the limiting factor for mobility in ultra-high-quality GaAs 2DESs \cite{100M1,100M2,DW1}. Efforts to systematically purify the Ga and Al source materials are in line with this understanding, as these metals are the most likely origin of unwanted impurities in a well-maintained, ultra-high-vacuum growth chamber \cite{Ga1,Ga2,Al}. Despite the different history of the source materials loaded into the MBE chambers and the varying purification techniques, ultra-high-quality samples from several groups around the world seem to converge at roughly similar values in electron mobility \cite{PhysicaE,Umansky,mobilityrocks,Ga1}. These results hint that the impurities that are limiting the mobility in the modern ultra-high-quality GaAs 2DESs come from somewhere other than the source material.
	
	
	While the environment in ultra-high-vacuum MBE chambers is certainly very sparse in atomic/molecular density, it is not completely void of matter. Even in well-equipped, conventional vacuum chambers, it is common that the mass spectrometer data of the growth space show traces of H$_2$O, N$_2$, O$_2$, CH$_4$, and their derivatives. If incorporated during growth, these species would act as impurities in the structure and cause degradation in the quality of the GaAs 2DES. Once the source materials become so pure that they are no longer the primary supplier of impurities to the growth environment, the background vacuum quality would determine the impurity concentration in the sample. 
	
	As shown in Fig. 1(a), we built an MBE chamber to verify this hypothesis. Our design includes conventional MBE chamber components such as a load-lock chamber and a liquid N$_2$ shroud, as well as three auxiliary cryo-cooled ($\sim17$ K) cold plates that augment four large (3000 $l$/s) cryopumps to achieve extreme levels of vacuum during sample growth. The cold plates are made of Cu to maximize cooling power and are coated with Ni to prevent possible corrosion when exposed to As and Ga that are ubiquitously present in the growth chamber. It is difficult to quantitatively assess how much the base pressure improved in our growth chamber with the cold plates operating because the ion extractor gauges installed in the chamber cannot reliably measure pressures below $P\sim2-3\times10^{-12}$ Torr, and the base pressure already reaches this range even when only the four cryopumps are pumping on the chamber. However, when the cold plates are turned on, the mass spectrometer data shows a factor of 10 improvement in the partial pressures of N$_2$ and O$_2$ species and a factor of 2.5 improvement in the partial pressures of H$_2$O-related species. The mass spectrometer data, as well as more details concerning the cold plates can be found in Section I of the Supplementary Information.  
	
	
	As previously demonstrated, it is possible to systematically evaluate the cleanliness of the growth environment during the MBE of GaAs/AlGaAs heterostructures \cite{Al}. A brief summary of the concept is to use the mobility of a GaAs 2DES with the sample structure shown in Fig. 1(b) as a very sensitive probe to gauge impurity accumulation during growth. This method utilizes the fact that impurities surface-segregate on the growth front of the back AlGaAs layer and deposit at the AlGaAs/GaAs interface when the growing layer is changed from AlGaAs to GaAs. For a given back AlGaAs layer thickness, the mobility is lower when the growth environment is worse. The strength of this procedure is that the back AlGaAs layer thickness ($d$) can be made extremely large to detect even the most minute amounts of impurities incorporated in the crystal during the growth process.
	
	Figure 1(c) shows the mobility of such structures grown in varying growth environments as a function of $d$. The enhanced mobility of samples grown after the sources were sufficiently outgassed (blue symbols) compared to those grown with a fresh batch of source materials (black symbols) demonstrates the importance of preparing pure source materials when aiming for the cleanest growth conditions. Under these conditions, the source materials were clean enough that high-mobility GaAs 2DESs made from them displayed mobilities on par with ultra-high-quality samples in the literature. Despite further purification efforts, we could not obtain a significant improvement from the blue data set under the normal operating vacuum conditions of having four cryopumps operating. This implied that at this point, our source materials had been amply purified so that they were no longer the primary supplier of impurities for our samples. It is then plausible to assume that the background vacuum starts to play a more important role. Consistent with this assumption, samples grown with only one cryopump operating exhibited worse mobilities compared to those grown with all four pumps turned on. Following this test, we investigated a series of samples grown with all the cryopumps and the cold plates operating during growth (data shown in red in Fig. 1(c)). The data clearly reveal that there is a significant improvement in GaAs 2DES mobility at all values of $d$ compared to the case when the cold plates are off. Remarkably, the mobility of the samples grown with all cold plates operating sustains the high value of $\mu\simeq8\times10^6$ cm$^2$/Vs even when $d=10$ $\mu$m.
	
	
	
	These results clearly demonstrate that vacuum integrity plays a crucial role in determining the amount of unintentional impurities deposited on the sample during growth once the source material has been extensively purified. Consequently, we grew several GaAs samples with a wide range of 2D electron densities to investigate the impact of having an ultra-clean vacuum environment in the MBE chamber. Figure 2(a) compares the mobility of these samples with the previous ultra-high-quality GaAs 2DESs \cite{PhysicaE,Umansky,mobilityrocks,Ga1}. Figure 2(b) shows the layer structure of the samples used to obtain the data presented in Fig. 2(a). We used two types of structures, the standard modulation-doped structure and the doping-well structure (DWS) \cite{DW3}. While the donor energy level, which determines the position of the Fermi level, is tied to the AlGaAs barrier in the standard modulation-doped structure, in the DWS it is tied to the narrow AlAs layers that flank a narrow GaAs doping quantum well \cite{DW3}. The DWS is advantageous in comparison to the standard modulation-doped structure because the electrons confined to the AlAs layer in the doped region provide additional screening for the 2DES from both residual impurities and intentional dopant ions \cite{DW1,DW3}. It is striking that with the improvement in vacuum, even the standard modulation-doped samples (black circles in Fig. 2(a)) have significantly higher mobility values for all densities when compared to previous state-of-the-art DWSs (red open circles). This is particularly noteworthy considering that DWSs were necessary to achieve the previously reported ultra-high-mobility values \cite{PhysicaE,Umansky,mobilityrocks,Ga1}. It seems that the reduction in impurities from better vacuum conditions by implementing the cold plates is significant enough to overcome the lack of such screening in our standard modulation-doped structures.
	
	\begin{figure*}[t]
 
 \centering
    \includegraphics[width=.95\textwidth]{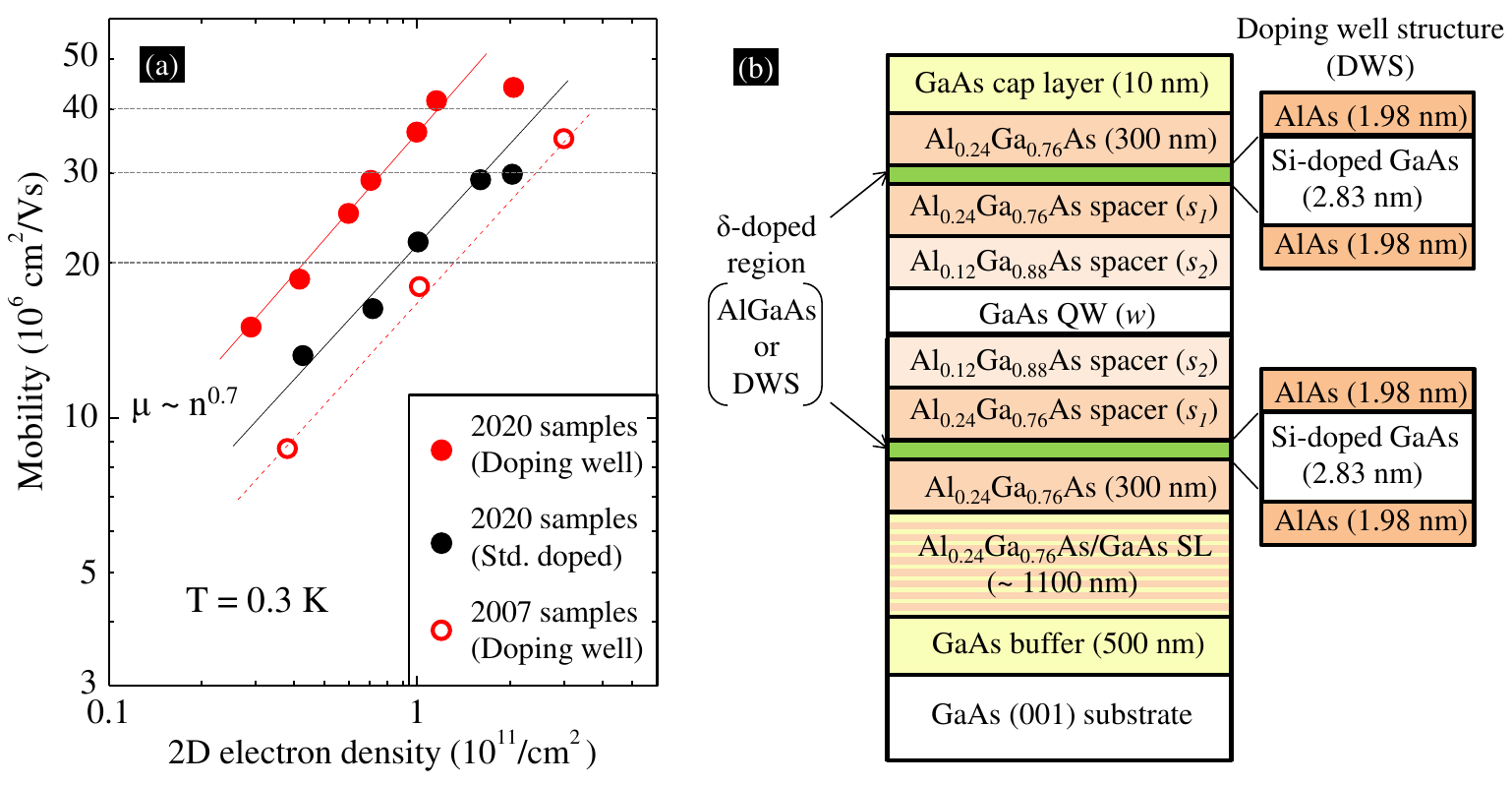} 

  \caption{\label{fig2} Mobility vs. 2D electron density for our GaAs 2DESs. (a) Data from samples grown under the improved vacuum conditions are shown as solid symbols for the doping-well (red) and standard modulation-doped (black) structures. The mobility values from our previous, ultra-high-quality, doping-well samples are shown as red open symbols for comparison. These previous samples have electron mobilities similar to ultra-high-quality samples reported in the literature \cite{PhysicaE,Umansky,mobilityrocks,Ga1}. The solid and dashed lines through each set of data points serve as guides to the eye, and their slopes are consistent with a power-law relation that is roughly $\mu\sim n^{0.7}$. (b) Layer structure of the samples used in (a). The spacer thicknesses and well widths of some representative samples can be found in Section II of the the Supplementary Information.}
\end{figure*}

	Furthermore, when we grow DWSs with the cold plates operating, we see an even larger increase in the mobility (red solid circles in Fig. 2(a)). These samples display mobility values as high as $\mu\simeq44\times10^6$ cm$^2$/Vs at the density of only $n\simeq2.0\times10^{11}$/cm$^2$. This implies a remarkable enhancement in sample quality considering that previous ultra-high-quality GaAs 2DESs had mobilities of $\mu\simeq35\times10^6$ cm$^2$/Vs at $n\simeq3.0\times10^{11}$/cm$^2$ \cite{PhysicaE,Umansky,mobilityrocks,Ga1}. Sample improvement is evident over a wide range of 2D electron densities as shown in Fig. 2(a). When $n<1.5\times10^{11}$/cm$^2$, our samples have mobility values that are roughly twice that of previous ultra-high-quality samples. For example, the mobility of our $n\simeq1.0\times10^{11}$/cm$^2$ sample is $\mu\simeq36\times10^6$ cm$^2$/Vs whereas previous ultra-high-mobility values for GaAs 2DESs with a similar density are less than $\mu\simeq18\times10^6$ cm$^2$/Vs \cite{Ga1,Pan2}.
	
\begin{figure*}[t]
 
 \centering
    \includegraphics[width=.95\textwidth]{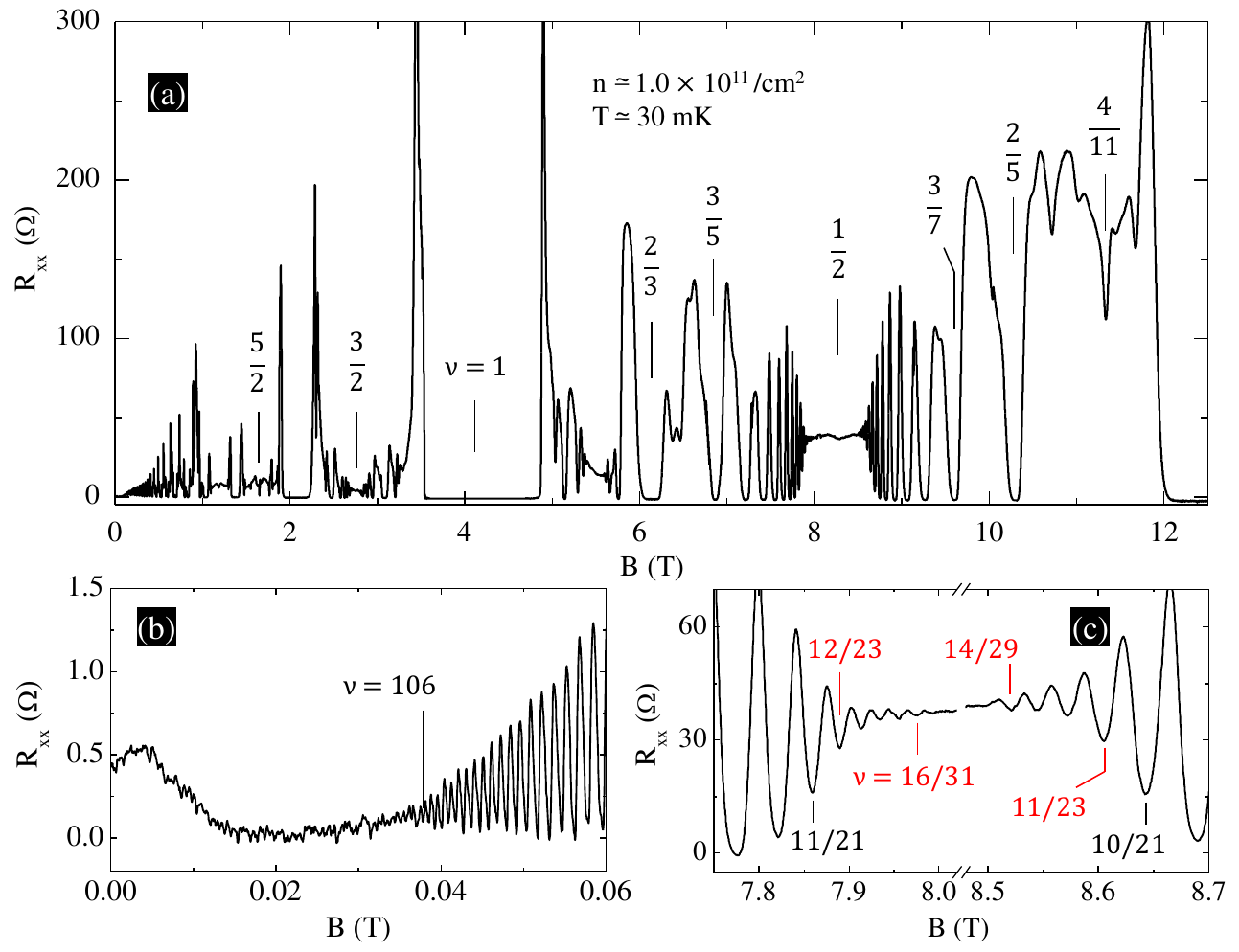} 

  \caption{\label{fig3} Low-temperature ($T\simeq30$ mK) magnetoresistance data of a GaAs 2DES with density $n\simeq1.0\times10^{11}$/cm$^2$. (a) A full-field $R_{xx}$ vs. magnetic field trace. The Landau level fillings ($\nu$) of several quantum Hall features are marked. (b) Expanded view of the low-field magnetoresistance near $B=0$. Resistance minima at fillings as high as $\nu=106$ can be seen in the Shubnikov-de Hass oscillations. (c) Expanded view of the magnetoresistance near $\nu=1/2$. High-order FQH states up to $\nu=11/21$ and 10/21 have been observed in previous ultra-high-quality samples with similar electron density (marked in black) \cite{Pan2}. New FQH states at $\nu=12/23$, 13/25, 14/27, 15/29, and 16/31 are seen on the left of $\nu=1/2$, and FQH states at $\nu=11/23$, 12/25, 13/27, and 14/29 on the right flank of $\nu=1/2$. The lowest- and highest-order new FQH state on each side of $\nu=1/2$ is marked in red.}
\end{figure*} 
	
	The power-law dependence observed for the mobility vs. 2D electron density profiles plotted in Fig. 2(a) is also noteworthy. Within the same category of samples, we observe a $\mu\propto n^{0.7}$ relation for all cases. A similar power-law dependence was reported in high-quality, low-density GaAs 2DESs with very large spacer-layer thicknesses, and it is usually interpreted as an indication that the mobility is limited by the residual impurities in the structure \cite{powerlaw,powerlaw2,powerlaw3}; there is also theoretical justification for such an interpretation \cite{Gold1}. This understanding is certainly consistent with our results, as the primary improvement we have made to our samples compared to the previous data is the reduction of residual impurities in the structure. Based on models for the two types of structures used for our samples \cite{DW1,Gold1}, we estimate that the residual impurity concentration in our GaAs QWs is $\simeq1\times10^{13}$ /cm$^3$. Considering that there are $\sim1\times10^{23}$ atoms/cm$^3$ in single-crystal GaAs, this means there is roughly one impurity for every 10 billion atoms in these samples.
	
		
	As mentioned in the introduction, probing intricate many-body phenomena is a core application of ultra-high-quality GaAs 2DESs. In this context, we also studied the low-temperature ($T\simeq30$ mK) magnetotransport of one of our representative samples with $n\simeq1.0\times10^{11}$/cm$^2$. The specific layer structure for this sample is provided in Supplementary Information Table 1. Figure 3(a) shows a full-field longitudinal magnetoresistance ($R_{xx}$) trace of the sample, while Figs. 3(b) and (c) focus on specific regions near zero magnetic field and near $\nu=1/2$, respectively; $\nu=hn/eB$ is the Landau level filling factor, where $h$ is the Planck constant, $e$ is the fundamental charge, and $B$ is the perpendicular magnetic field. It is clear from the data that the sample has very high quality. For example, as seen in Fig. 3(b), there are prominent signatures of Shubnikov-de Hass oscillations up to $\nu=106$ at $B<0.04$ T. This implies that the Landau level broadening in this sample is smaller than the $B=0.04$ T cyclotron energy gap of $heB/m^*\simeq68$ $\mu$eV ($m^*=0.067$ is the effective mass of electrons in GaAs in units of the free-electron mass). In addition, the data plotted in Fig. 3(c) display high-order FQH states up to $\nu=16/31$ and $\nu=14/29$ on the left and right flanks of $\nu=1/2$. We compare these results to those reported previously for epitaxially-grown samples and two-dimensional materials with ultra-high-quality. In total, compared to previous ultra-high-quality GaAs samples with similar density \cite{Pan2}, nine extra FQH states are observed near $\nu=1/2$ in our sample, whose lowest- and highest-order Landau level fillings are marked in red on each side of $\nu=1/2$. For comparison, in ultra-high-quality monolayer graphene samples, high-order FQH states have been observed up to $\nu=8/15$ and $\nu=7/15$ on the left and right flanks of $\nu=1/2$ at similar temperatures but higher magnetic fields ($\sim14$ T) \cite{Graphene.FQHE.2018}; the data presented in Fig. 3 exhibits 15 additional FQH states with respect to these samples.
	

\begin{figure*}[t]
 
 \centering
    \includegraphics[width=.95\textwidth]{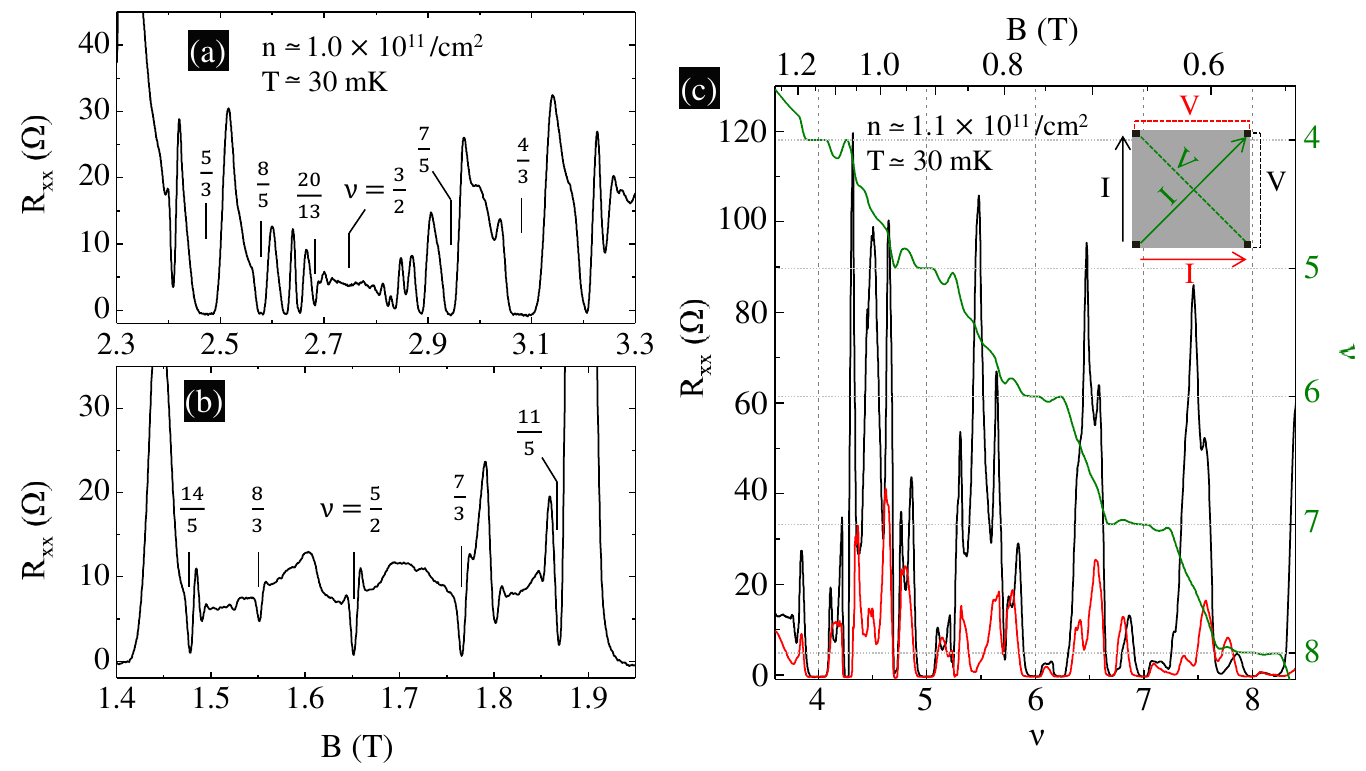} 

  \caption{\label{fig4} Low-temperature ($T\simeq30$ mK) magnetotransport data of $n\simeq1.0\times10^{11}$/cm$^2$ samples at higher Landau level fillings. (a) Magnetoresistance data near $\nu=3/2$ ($N=0$ orbital Landau level) and (b) $\nu=5/2$ ($N=1$). The number of observed FQH states as well as the strengths of each of the features are remarkable, considering the relatively low density of the sample. (c) Longitudinal (black and red) and Hall (green) resistance data for the high-index ($N=2$ and 3) orbital Landau levels. The presence of reentrant integer quantum Hall states and stripe/nematic phases are evident at magnetic fields $B<1$ T. The inset shows the configuration of current ($I$) and voltage ($V$) contacts used for the measurements.}
\end{figure*} 
	
	The quality of our samples in this density range also stand out at higher Landau level fillings. Figures 4(a) and (b) show expanded $R_{xx}$ traces of Fig. 3 sample near $\nu=3/2$ and $\nu=5/2$, respectively. Remarkably, even at this relatively low density, FQH states up to $\nu=20/13$ are observable in the vicinity of $\nu=3/2$. Furthermore, the $\nu=5/2$ and other FQH states in the second orbital Landau level ($N=1$) are extraordinarily strong, considering that they occur at $B<1.9$ T. In fact, the activation gap we measure for the $\nu=5/2$ FQH state is $^{5/2}\Delta\simeq820$ mK (see Supplementary Information Fig. S4), surpassing the value of $^{5/2}\Delta\simeq625$ mK measured in previous ultra-high-quality samples by a significant margin \cite{Manfrafivehalf}. This observation is particularly noteworthy considering the fact that the density of this sample is only $\sim1/3$ of the 2DESs used in previous evaluations. 
	
	Given the potentially non-Abelian nature of the $\nu=5/2$ FQH state \cite{Nayak,WillettPNAS}, the data presented here have exciting implications for the realization of fault-tolerant, topological quantum computing devices. Using optimistic estimates for qubit error rate \cite{Nayak}, our larger $^{5/2}\Delta$ value implies a factor of $\sim10^6$ improvement compared to previous ultra-high-quality samples when operations are performed at $T=5$ mK. In principle, if the samples presented here do not suffer from severe detrimental effects caused by lithographic procedures and are amenable to gating, one should be able to perform significantly more robust qubit operations at much lower magnetic fields. Previously, samples with similar structure have been used for interferometry experiments that can be considered a basis for qubit operation in GaAs 2DESs \cite{WillettPNAS}, so we are optimistic that the technical details can be worked out.

	Figure 4(c) shows magnetotransport data of a different sample with a similar density of $n\simeq1.1\times10^{11}$/cm$^2$ in magnetic field ranges that correspond to higher ($N=2$ and 3) orbital Landau levels. Well-quantized, reentrant integer quantum Hall states, as well as signatures of stripe/nematic phases, are observable at magnetic fields $B\lesssim1.0$ T. These correlated states are known to be fragile, and are typically only observed in much higher density samples at larger magnetic fields and lower temperatures \cite{stripe1,stripe2,Friess.Nat.Phys.2017,RoBubble}.
	
	The results presented here suggest a bright future for the investigation of interaction-driven physics in GaAs 2DESs. With this improvement in sample quality, several new FQH states have emerged and numerous correlated phases display strong robustness. For example, the $\nu=5/2$ FQH state exhibits a gap value of $^{5/2}\Delta\simeq820$ mK, and we demonstrate that reentrant integer quantum Hall and stripe/nematic states are clearly visible even at very low magnetic fields and electron densities. Moreover, we have experimentally shown that vacuum integrity limits sample quality in current state-of-the-art MBE-grown GaAs. This gives a clear direction for further improvement in the quality of GaAs 2DESs. We speculate that further reducing the residual background impurity concentration by a factor of $\sim3$ should allow one to obtain mobility values exceeding $10^8$ cm$^2$/Vs as predicted by theory \cite{100M1}. Perhaps this can be realized by implementing even more cold plates or pumps in the MBE chamber. 
	
	Some mysteries have also developed. When $n\geq1.5\times10^{11}$/cm$^2$, the electron mobilities seem to deviate to lower values than the power-law relation $\mu\propto n^{0.7}$ would predict. We are currently unsure of the origin of this behavior. It is possible that the remote ionized impurities from the intentional dopant atoms become relevant as a smaller spacer thickness is required to achieve higher 2D electron densities. If this is the case, in the future it may be useful to start from a low density sample with large spacer thickness and increase the  density by applying gate voltages to circumvent this issue. Another option is to vary structural parameters in the doped region of the DWS so that the spacer thickness can be increased while maintaining the same 2DES density in the main quantum well \cite{DW3}.

\newpage

{\bf Methods}

{\bf Sample preparation}

All of our samples are grown on 2-inch-diameter GaAs substrates in the vacuum chamber setup shown in Fig. 1(a). The substrates are outgassed for 30 minutes at $T\simeq610$ \textdegree C in an As beam flux of $P\sim6.0\times10^{-6}$ Torr prior to growth, where the temperature of the substrate was evaluated using a factory-calibrated pyrometer (Ircon Modline 7). We always confirm clear single-crystalline features in the reflection-high-energy-electron-diffraction (RHEED) patterns of the substrate after this process. The substrate temperature is typically $T\simeq640$ \textdegree C during growth. This growth temperature was chosen based on our past experience for growing ultra-high-quality samples. For such samples, the typical temperature window for optimal growth was $\simeq\pm10$ \textdegree C. The deposition rate of GaAs is calibrated to be $\simeq2.83$ {\AA}/s for all growths by tuning the temperature of the Ga oven based on RHEED oscillations. We tune the Al growth rate in a similar fashion to obtain the barrier alloy fraction of choice. The barrier alloy fraction is 32\% for the samples whose data are shown in Fig. 1(c) while the ultra-high-mobility samples use a stepped-barrier structure with alloy fractions 24\% and 12\%. The specifics of the sample structure of the ultra-high-mobility samples can be found in Supplementary Information Section II.

{\bf Transport measurements}

We performed all electronic measurements in the van der Pauw configuration using low-frequency lock-in amplifiers. Our samples have a square shape and a typical size of 4 mm$\times4$ mm. The mobility values of the GaAs 2DESs are evaluated in a $^3$He cryostat with a base temperature of $T\simeq0.3$ K. A simple Drude formula $\mu=1/\rho ne$ is used to obtain the mobility, where $n$ is the 2DES density, $e$ is the fundamental electron charge, and $\rho$ is the resistivity of the 2DES. Quantum Hall features in the magnetoresistance data are used to deduce $n$. For $\rho$, we take the average value of the resistance ($R_{ave}$) measured between all the four-probe contact configurations in the sample, and use the standard van der Pauw geometry expression $\rho=\pi R_{ave}/ln(2)$. The low-temperature magnetotransport data presented in the main text are measured in a dilution refrigerator with a base temperature of $T\simeq30$ mK. It is well known that the illumination and cooldown procedure can affect the electronic properties of the GaAs 2DES at low temperatures. For the measurements in this work, we illuminate all samples for 5 minutes at $T\simeq10$ K with a red light-emitting diode (LED) before turning the LED off and waiting for an additional 30 minutes at $T\simeq10$ K. A current of 6 mA is passed through the LED during illumination. We then cool the samples down to the base temperature. A similar procedure is adopted for the dilution refrigerator measurements but the LED illumination is done at $T\simeq4$ K instead of 10 K because of the limitations of the apparatus. By repeating this procedure, we saw less than 5\% variance in the 2DES density and mobility even when the samples experienced a full thermal cycle to room temperature. For the magnetotransport measurements, a magnetic field sweep rate of 1 T/hour was used except for the case of Fig. 1(b), where a slower sweep rate of 0.1 T/hour was used.

\begin{acknowledgments}
We acknowledge support through the NSF (Grants DMR 1709076 and ECCS 1906253) for measurements, and the NSF (Grant MRSEC DMR 1420541), the Gordon and Betty Moore Foundation's EPiQS program (Grant GBMF9615 to L. N. P.), and the Department of Energy (DOE) Basic Energy Sciences (Grant DE-FG02-00-ER45841) for sample fabrication and characterization.
 \end{acknowledgments}

 {\bf Data availability} \par
 Source data are provided with this paper. Data supporting the results in this paper and the supplementary material is available on request to the corresponding author.
 
 {\bf Author contributions} \par
 Y. J. C. and L. N. P. conceived the work. K. W. B., K. W. W., and L. N. P. designed and built the molecular beam epitaxy chamber. Y. J. C., K. W. B., K. W. W., and L. N. P. designed, grew, and evaluated the quality of all samples at $T\simeq0.3$ K. K. A. V. R. and P. T. M. performed dilution refrigerator measurements. Y. J. C. and M. S. wrote the manuscript with input from all co-authors.
 
  {\bf Competing Interests} \par
The authors declare no competing interests.

 \textsuperscript{$\dag$} These authors contributed equally to this work.

\end{document}


\setcounter{page}{1}
\setstcolor{red}

\renewcommand{\thetable}{S\arabic{table}}  
\renewcommand{\thefigure}{S\arabic{figure}}

\title[]{Supplementary Information for ``Ultra-high quality two-dimensional electron systems"}
\author{Yoon Jang \surname{Chung}\textsuperscript{$\dag$}}
\author{K. A. \surname{Villegas-Rosales}\textsuperscript{$\dag$}}
\author{K. W. \surname{Baldwin}}
\author{P. T. \surname{Madathil}}
\author{K. W. \surname{West}}
\author{M. \surname{Shayegan}}
\author{L. N. \surname{Pfeiffer}}
\affiliation{Department of Electrical Engineering, Princeton University, Princeton, NJ 08544, USA  }
\date{\today}

\maketitle

\section{\bf{I.} Details on the 17 K cold plates}

\begin{figure*}[h]
\centering
    \includegraphics[width=.80\textwidth]{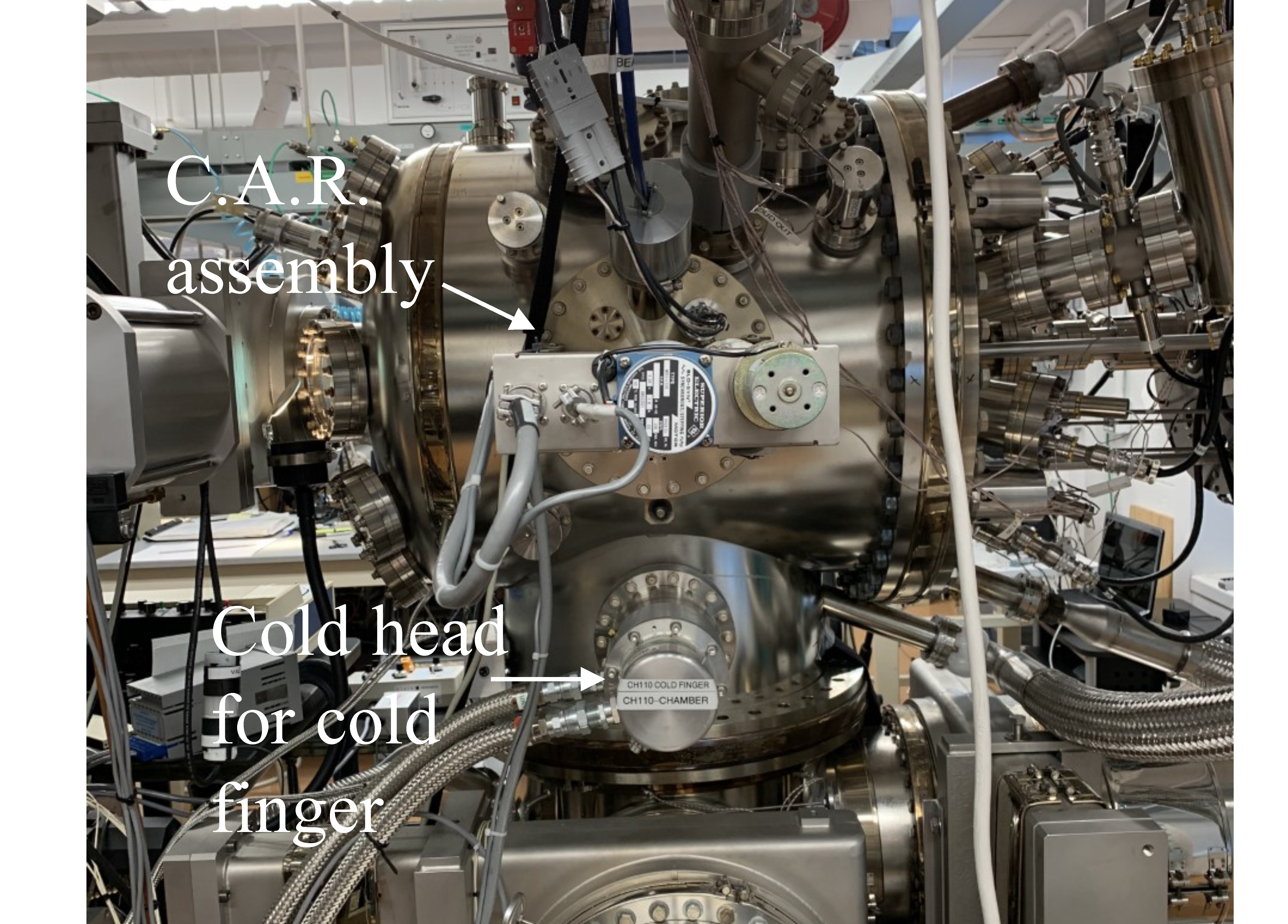}

 \caption{\label{figS1}Photograph of the exterior of the growth chamber. The cold head for the cold plate in the vicinity of the growth space, or the `cold finger', is installed on the growth chamber via an auxiliary port beneath the port that the continual azimuthal rotation (C.A.R.) assembly is installed. The cold plates in the sump have cold heads installed on the auxiliary ports in the sump, as schematically shown in Fig. 1(a) of the main text.}
\end{figure*}

Each cold plate is cooled by a cold head that is installed on the growth space or sump through auxiliary ports in the vacuum chamber, as shown in Figs. 1(a) and S1. All three single-stage cold heads have a cooling power of 180 W at 77 K (custom modified version of the Sumitomo CH-110). The Ni-coated Cu cold plates are bolted to the Cu face of the cold heads with an In washer between them to ensure good thermal contact. Three Si diodes (Lakeshore DT-470) are attached to each assembly to monitor the temperature during operation. One diode is installed on the Cu face of the cold head, while another is installed on the other side of the Ni-coated Cu cold plate near the center axis of the cold head to evaluate whether the cold plate is thermally equilibrated with the cold head. The final diode is installed close to the upper edge of the cold plate to check the uniformity of temperature across the cold plate. When first turned on, after reaching base temperature, all thermometer readings yielded temperatures close to 17 K. The three diodes on each cold plate typically read values within 1 K of each other.

Once the cold plates are turned on and reach base temperature, we see a drastic improvement in the mass spectrometer data of the vacuum chamber, as shown in Fig. S2. The N and O related peaks at mass/charge ratios of 14,16, and 28 that were clearly discernible from the background with the cold plates off virtually vanish when the cold plates are operating. The series of peaks associated with H$_2$O also show a significant decrease in intensity when the cold plates are on.

\begin{figure*}[t]
\centering
    \includegraphics[width=.55\textwidth]{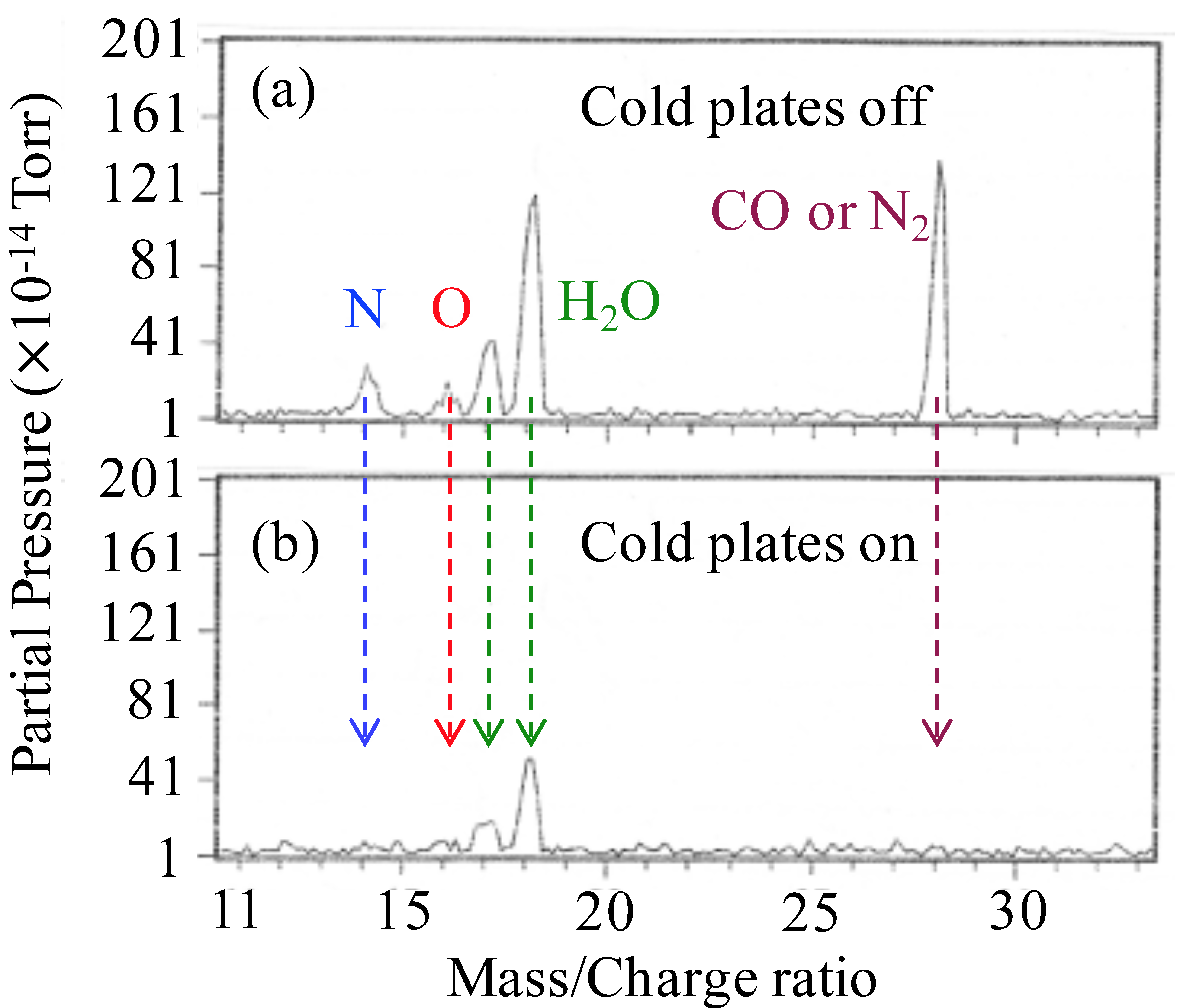}

 \caption{\label{figS2}Mass spectrometry data of the growth chamber (a) before and (b) after the 17 K cold plates are turned on. As discussed in the main text, the partial pressures of N, O, H$_2$O, and CO or N$_2$ species show a significant decrease with the operation of the cold plates.}
\end{figure*}

Preliminary results suggest that we can grow samples with mobility values comparable to the best samples shown in Fig. 2(a) even when only the two larger cold plates in the sump are operating. It is possible that the much ($\sim14$ times) larger surface area and hence pumping efficiency of these cold plates compared to the one located in the vicinity of the growth space (the sample growth shroud) is responsible for this behavior. However, at this time, we can not fully rule out the potentially positive influence of the smaller cold plate in the vicinity of the growth space on high-quality sample growth.

\newpage   

\section{\bf{II.} Sample structure for the ultra-high-quality two-dimensional electron systems}

\begin{figure*}[h]
\centering
    \includegraphics[width=.80\textwidth]{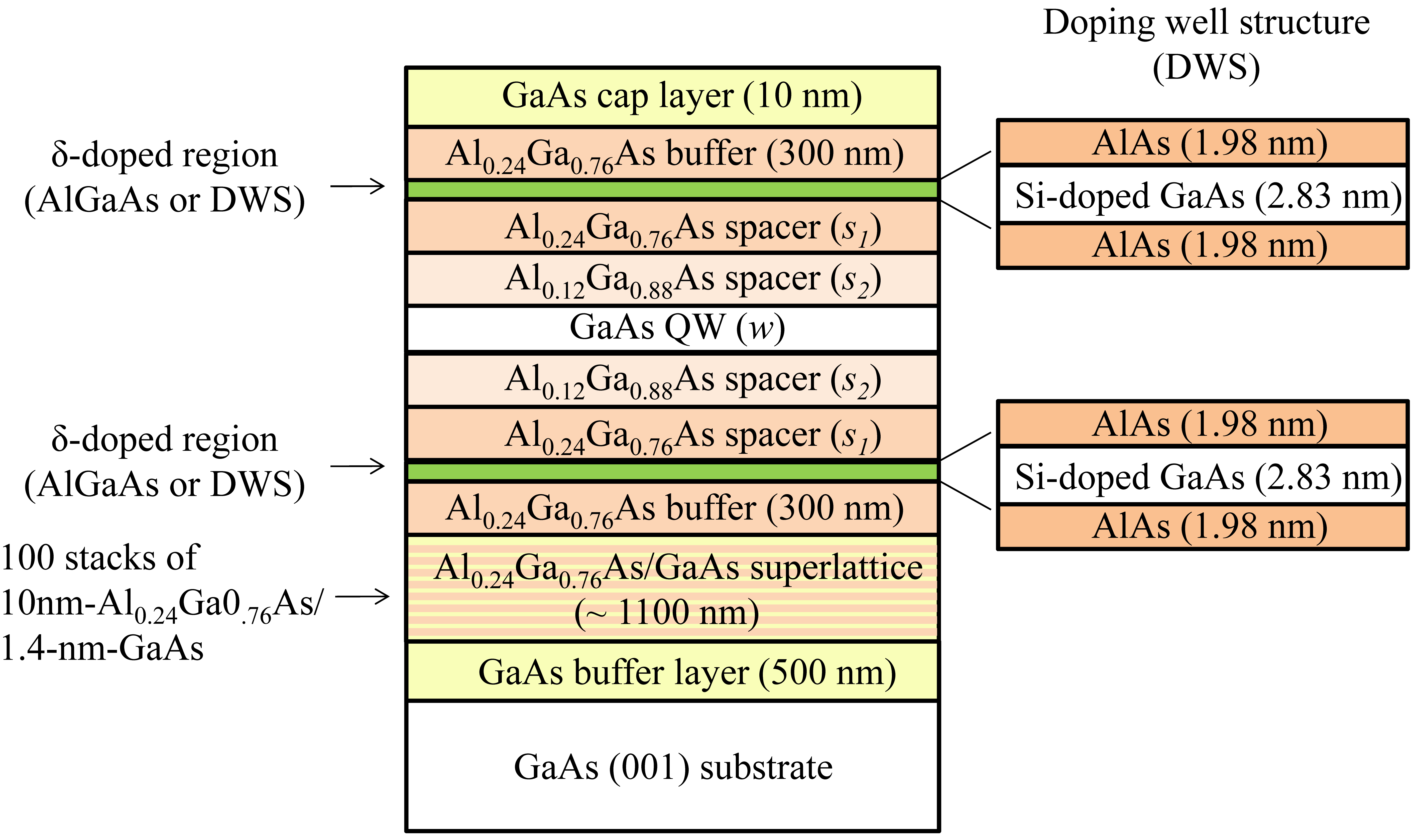}

 \caption{\label{figS3}Sample structure for the ultra-high-quality GaAs two-dimensional electron systems whose data are shown in the main text. The spacer layer thickness is varied to obtain the desired sample density, and the quantum well width is chosen accordingly to prevent second-subband occupation.}
\end{figure*}

All ultra-high-quality samples follow the structure shown in Fig. S3. Standard modulation-doped structures are $\delta$-doped in AlGaAs while the doping-well structures (DWSs) use the doping scheme described in the right panels of the figure. A 12\%/24\% stepped-barrier structure is implemented to reduce the Al composition of the barrier directly in contact with the main quantum well. The thickness of each of the barrier layers is controlled so that no parallel channel forms at the 12\%/24\% barrier interface, and the total spacer thickness is varied to attain the desired sample density. The well width of the quantum well is varied for each sample so that there is no second-subband occupation. Specific structural parameters are summarized in Table S1 for some representative samples.

\newpage

\begin{table}[h]
\begin{center}
\begin{tabular}{ |c|c|c|c|c| } 

\hline
 $n$ ($\times10^{11}$ /cm$^2$)& $\mu$ ($\times10^6$ cm$^{2}$/Vs) & $s_1$ (nm) &  $s_2$ (nm) & $w$ (nm)  \\
\hline
\hline
2.06 & 44.0 & 68.2 & 60.0 & 34.0 \\ 
\hline
1.15 & 41.5 & 114 & 100 & 45.3 \\ 
\hline
1.00 & 36.0 & 136 & 120 & 50.0 \\ 
\hline
0.71 & 29.0 & 195 & 171 & 58.5 \\
\hline
\end{tabular}

\caption{Structural parameters of some representative ultra-high-quality doping-well-structure samples whose data are shown in the main text; for definitions of $s_1$, $s_2$, and $w$, see the sample structure in Fig. S3.} \label{tab:table1}

\end{center}
\end{table}

\newpage

\section{\bf{III.} Measurement of energy gap for the $\nu=5/2$ fractional quantum Hall state}

\begin{figure}[h]
\centering
    \includegraphics[width=.50\textwidth]{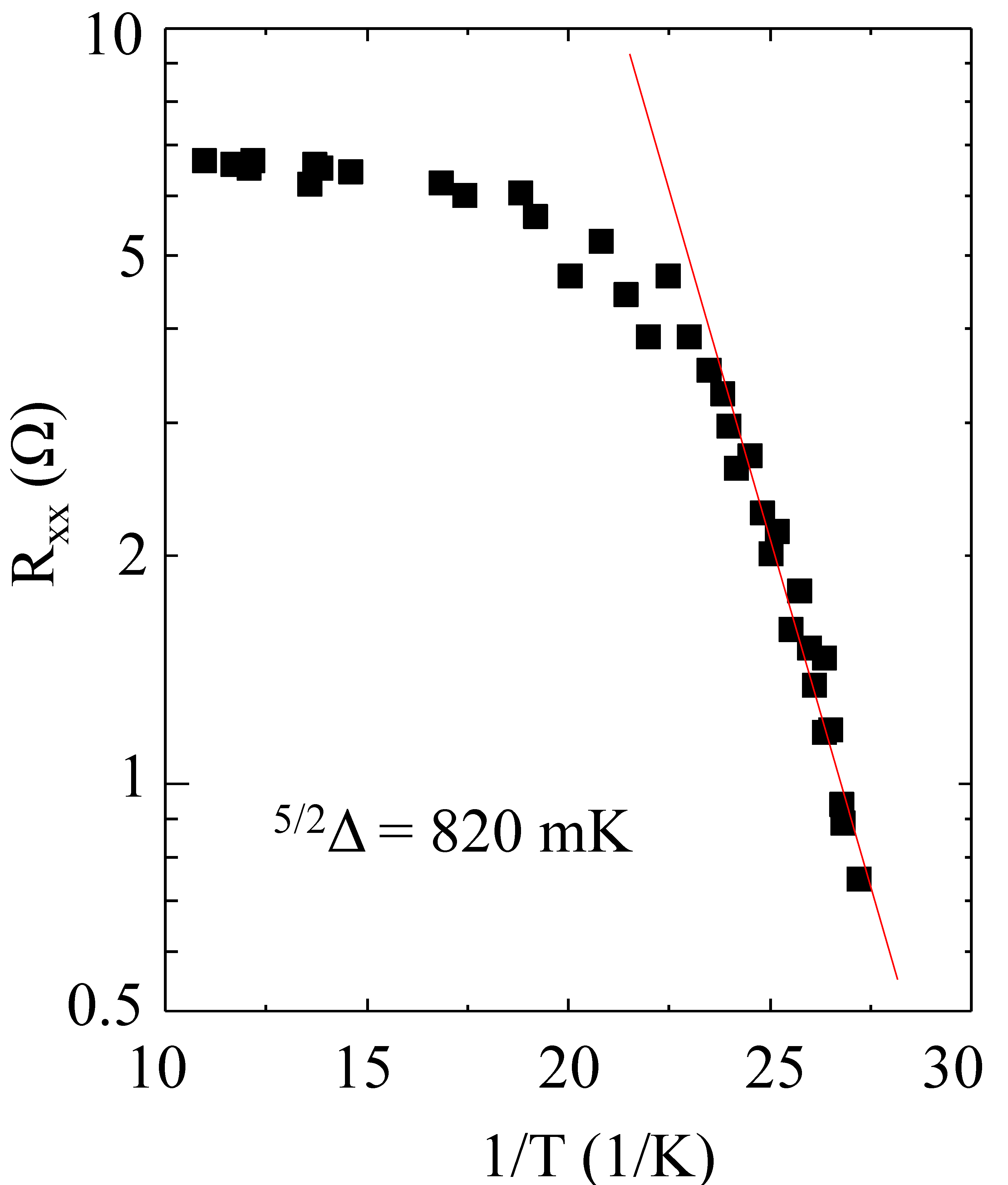}

 \caption{\label{figS4} Arrhenius plot of resistance at $\nu=5/2$. The sample density is $n\simeq1.0\times10^{11}$/cm$^2$. }
\end{figure}

Magnetoresistance traces were taken at a sweep rate of 0.1 T/hour in the vicinity of the $\nu=5/2$ fractional quantum Hall state to precisely determine the magnetic field position of the $R_{xx}$ minimum at base temperature. Then the energy gap $^{5/2}\Delta$ was measured by raising the temperature of the sample while monitoring the sample resistance at the magnetic field corresponding to $\nu=5/2$. The data are shown in the Arrhenius plot of Fig. S4. A line corresponding to $R_{xx}\sim e^{-^{5/2}\Delta/{2k_BT}}$ was fitted to the data plotted in Fig. S4 (shown in red), yielding an energy gap of $^{5/2}\Delta=820$ mK. Here $k_B$ is the Boltzmann constant.

\newpage




